\documentclass[useAMS,usenatbib]{mn2e}
\usepackage{graphicx}
\usepackage[utf8]{inputenc}
\usepackage{txfonts}

\title[SAX J2103.5+4545 in quiescence]{The quiescent state of the accreting X-ray pulsar 
SAX J2103.5+4545}

\author[P. Reig et al.]{P. Reig$^{1,2}$\thanks{E-mail: pau@physics.uoc.gr},
	 V. Doroshenko$^{3}$ and A. Zezas$^{2,1}$ 
\\
$^{1}$IESL, Foundation for Research and Technology, 71110 Heraklion,
Crete, Greece\\
$^{2}$University of Crete, Physics Department, PO Box 2208, 710 03
Heraklion, Crete, Greece \\
$^{3}$Institut für Astronomie und Astrophysik, Sand 1, 72076 Tübingen, 
Germany}

\newcommand{\sax}  {SAX\,J2103.5+4545}
\newcommand{\ha}   {H$\alpha$}

\newcommand{\ew}   {EW(H$\alpha$)}
\newcommand{\ergs}   {erg s$^{-1}$}

\def\simless{\mathbin{\lower 3pt\hbox
     {$\rlap{\raise 5pt\hbox{$\char'074$}}\mathchar"7218$}}}   
\def\simmore{\mathbin{\lower 3pt\hbox
     {$\rlap{\raise 5pt\hbox{$\char'076$}}\mathchar"7218$}}}   

\def\msun{~{\rm M}_\odot}

\def\ergs{erg s$^{-1}$}

\begin{document}

\date{Accepted ??. Received ??; in original form ??}

\pagerange{\pageref{firstpage}--\pageref{lastpage}} \pubyear{2014}

\maketitle

\label{firstpage}

\begin{abstract}

We present an X-ray timing and spectral analysis of the Be/X-ray binary
SAX\,J2103.5+4545 at a time when the Be star's circumstellar disk had
disappeared and thus the main reservoir of material available for accretion
had extinguished.  In this very low optical state, pulsed X-ray emission
was detected at a level of $L_X\sim 10^{33}$ \ergs. This is the lowest
luminosity at which pulsations have ever been detected in an accreting
pulsar.  The derived spin period is 351.13 s, consistent with previous
observations. The source continues its overall long-term spin-up, which
reduced the spin period by 7.5 s since its discovery in 1997. The X-ray
emission is consistent with a purely thermal spectrum, represented by a
blackbody with $kT=1$ keV. We discuss possible scenarios to explain the
observed quiescent luminosity and conclude that the most likely mechanism
is direct emission resulting from the cooling of the polar caps, heated
either during the most recent outburst or via intermittent accretion in
quiescence. 

\end{abstract}


\section{Introduction}

\sax\ belongs to the sub-class of high-mass X-ray binaries known as
Be/X-ray binaries (BeXB). In these systems, a neutron star orbits
around an OBe companion \citep{reig11}. In a BeXB, the main source of
matter available for accretion is the gaseous geometrically thin
equatorial disk around the Be star. The disk is fed from the material
lifted from the star's photosphere by a still uncertain mechanism. In
classical (isolated) Be stars, there is growing evidence that the
disk is Keplerian and supported by viscosity \citep{rivinius13a}.

Most BeXBs are transient X-ray sources that exhibit outbursts when a
compact object passes close or through the Be disk. Correlated
optical/IR/X-ray variability is often observed on time scales of months or
years and generally attributed to the extension of the circumstellar disk
\citep{negueruela98a,reig07b,reig10a}. The outburst activity is commonly
divided in two types: type I outbursts are modulated by the orbital period
of the system and occur when the neutron star passes close to the disk and
accretes from its outer regions. The type II, or giant outbursts, exhibit
higher X-ray luminosity close to the Eddington value $L_X\sim 10^{38}$ erg
s$^{-1}$, and are usually associated with the accretion of a substantial
part of the Be disk \citep{reig07b}.

Several systems have also been observed in quiescence at X-ray
luminosities in the range of $L_X\sim 10^{32}-10^{34}$ erg s$^{-1}$
\citep{schulz95,campana02,tomsick11}. However, pulsations were detected
only in the brighter sources with longer spin periods, which are likely
powered by accretion in quiescence
\citep{mereghetti87,negueruela00,rutledge07}.

The compact object in \sax\ is clearly a neutron star as the observed X-ray
emission is pulsed. At the time of its discovery by {\it BeppoSAX} in
February 1997, the pulse period was $P_{\rm spin}=358.61\pm0.03$ s
\citep{hulleman98}. Since then, the neutron star exhibits a general spin-up
trend, although the rate of the period change has not remained constant.
Occasionally, the long-term spin-up trend is interrupted by spin-down
intervals \citep{ducci08}. The optical companion is a moderately reddened
($A_V=4.2$ mag) V=14.2 B0Ve star \citep{reig04}.

The distance estimated from optical data is $\sim$6.5 kpc
\citep{reig04,reig10a}, while X-ray observations suggest a lower
value of $\sim$4.5 kpc \citep{baykal07}. \sax\ has a moderately
eccentric orbit with $e=0.4$ and an orbital period of $12.7$ days
\citep{baykal07,camero07}. Its relatively long spin period and
relatively short orbital period locates \sax\ in the wind-fed
supergiant region of the $P_{\rm orb}-P_{\rm spin}$ diagram
\citep{corbet86}.

\sax\ shows extended bright and faint X-ray states that last for several
months \citep{reig10a}. During the faint state, the X-ray intensity does
not change significantly with orbital phase \citep{baykal02,blay04}, and
the spin frequency of the neutron star remains fairly constant
\citep{baykal07} or slightly decreases \citep{ducci08}. In this state the
B-type companion shows \ha\ in absorption \citep{reig10a,kiziloglu09},
which signifies the recession of the circumstellar Be disk. The average
X-ray luminosity in this state is $L_X\sim 3\times 10^{35}$ erg s$^{-1}$ in
the 3--30 keV range (assuming a distance of 6.5 kpc).
The bright state generally starts with a sharp flare that lasts for
one to two orbital cycles. This flare is then followed by a
progressive increase in the X-ray intensity until a maximum is
reached at about one order of magnitude brighter than in the faint
state. During the bright states, the neutron star spins up
\citep{baykal07,camero07}, shows moderate outbursts modulated by the
orbital period \citep{baykal00,sidoli05}, and displays \ha\ line in
emission \citep{reig10a,kiziloglu09}, which indicates the growth
of the Be disk. The X-ray luminosity at the peak of the outbursts
ranges between $(0.6-1.0) \times 10^{37}$ erg s$^{-1}$, while at the
peak of the flare the luminosity is typically a factor of 2 higher.

The X-ray spectra of the bright state are distinctly harder than those of
the faint state \citep{baykal02,reig10a}. The 1--150 keV X-ray spectra are
well represented by an absorbed ($N_{\rm H}\approx \times 10^{22}$
cm$^{-2}$) power law ($\Gamma_{\rm bright}\approx 0.8-1$, $\Gamma_{\rm
faint}\approx 1.2-1.4$) plus and exponential cutoff at high energy ($E_{\rm
cut}\approx 13-18$ keV). In addition, a cool iron fluorescence line is observed at 6.4\,keV. At lower energy, a soft
component consistent with blackbody emission ($kT\approx1.9$ keV) has been
shown to be significant in an {\it XMM-Newton} observation during a bright
state \citep{inam04}. This observation also revealed a 44 mHz
quasi-periodic oscillation.

In this work we present the results of a timing and spectral analysis of a
{\it Chandra} observation aimed specifically to explore the characteristics
of the X-ray emission at very low accretion rates.   The source is expected
to be in X-ray quiescence when the material in the disk dissipates.  Thanks
to our regular monitoring of the evolution of the \ha\ line in the optical
spectrum of \sax, we were able to trigger the X-ray observations when the
line appeared in absorption. The source indeed turned out to be in deep
X-ray quiescence with a flux more than two orders of magnitude lower than
the faint state previously reported in {\it RXTE} observations and almost
four orders of magnitude lower than during the bright state. We discuss
the possible origin of the observed X-ray quiescent emission.

\begin{figure}
\resizebox{\hsize}{!}{\includegraphics{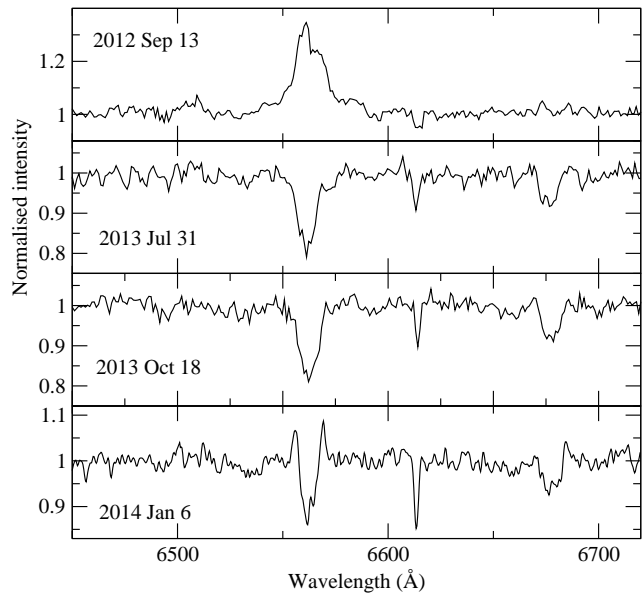} } 
\caption[]{Profile of the \ha\ line at different epochs. The {\it
Chandra} observations took place on September 9, 2013. }
\label{halpha}
\end{figure}

\begin{table}
\begin{center}
\caption{H$\alpha$ equivalent width measurements ($1\sigma$ errors). }
\label{ewha}
\begin{tabular}{cccc}
\hline \hline \noalign{\smallskip}  
Date	&Julian date	&EW(\ha)	 &Telescope	 \\
	&(2,400,000+)	&(\AA)		 &		 \\

\hline \noalign{\smallskip}
\multicolumn{4}{c}{\sax} \\
\hline \noalign{\smallskip}
13-09-2012 &56184.44	&$-6.4\pm0.4$	&SKO	\\
31-07-2013 &56505.39	&$+2.4\pm0.3$	&SKO	\\
30-08-2013 &56535.36	&$+2.3\pm0.2$	&SKO 	\\
18-10-2013 &56584.27	&$+1.9\pm0.1$	&SKO	\\
03-11-2013 &56599.62    &$+1.9\pm0.1$	&FLWO	\\
07-12-2013 &56634.59    &$+1.8\pm0.1$	&FLWO	\\
06-01-2014 &56664.57    &$+0.35\pm0.05$	&FLWO	\\
\hline \noalign{\smallskip}
\end{tabular}
\end{center}
\end{table}

\section{Observations}

\subsection{Optical observations and the \ha\ line profile}

Emission lines in Be stars are the result of recombination radiation from
ionised hydrogen in the hot, extended circumstellar envelope surrounding
the central Be star. The \ha\ line is the prime indicator of the
circumstellar disk state. In particular, its equivalent width (\ew) is a
robust tracer of the size of the disk
\citep{quirrenbach97,tycner05,grundstrom06}. In the absence of the disk, no
emission is expected and the line should have the typical photospheric
absorption profile. 

In this section we present optical spectroscopic observations that
demonstrate the absence of the equatorial disk at the time of the X-ray
observations. The optical spectroscopic observations were obtained from the
Skinakas observatory (SKO) in Crete (Greece) and from the Fred Lawrence
Whipple observatory (FLWO) at Mt. Hopkins (Arizona). Table~\ref{ewha} gives
the log of the spectroscopic observations and the measured H$\alpha$
equivalent width. The 1.3\,m telescope of the Skinakas Observatory was
equipped with a 2000$\times$800 (15 $\mu$m) pixel ISA SITe CCD and a 1302
l~mm$^{-1}$ grating, giving a nominal dispersion of $\sim$1 \AA/pixel. The
FLWO observations were carried out in queue mode with the 1.5-m telescope
equipped with the FAST-II spectrograph \citep{fabricant98} and the 1200
l~mm$^{-1}$ grating, yielding a dispersion of 0.4\AA/pixel. The data were
analysed using the RoadRunner package \citep{tokarz97} implemented in IRAF.
Spectra of comparison lamps were taken before each exposure in order to
account for small variations of the wavelength calibration during the
night.

Our monitoring of \sax\ reveals that the \ha\ profile changed from emission
into absorption some time around March-May 2013. By convention, the
equivalent widths of absorption lines are expressed as positive numbers,
while the equivalent widths of emission lines are quoted as negative. 
Fig.~\ref{halpha} shows the profile of the \ha\ line at different epochs.
In September 2012, a strong emission asymmetric \ha\ profile with an
equivalent width $EW(H\alpha)=-6.4$ \AA\ was measured, indicating the
presence of the circumstellar disk. The disk also affected the shape of the
HeI line at 6678 \AA, which appeared with a double peaked emission profile.
About a year later, both lines were observed in absorption, although a very
weak peak emerging from the center of the  line was observed. This type of
profile is known as central quasi-emission peak (CQE) and is an indication
of a highly debilitated disk \citep[see][and references therein]{reig10a}.
On January 6, 2014, the line displayed a double peak shell profile, which
implies that the disk is reforming again. The shell feature (central peak
below the continuum) suggests that we see the disk at high inclination
angle.

\begin{figure}
\resizebox{\hsize}{!}{\includegraphics{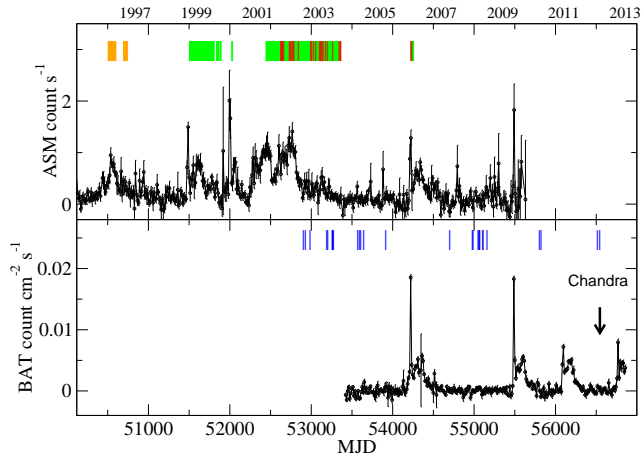} } 
\caption[]{Long-term X-ray variability of \sax\ as seen by the all-sky
monitors {\it RXTE}/ASM (top) and {\it SWIFT}/BAT light curves (bottom).
The original 1-day resolution light curves were rebinned to a bin size 
equal to the orbital period ($P_{\rm orb}=12.67$ d). The time of published 
 X-ray observations of various missions is indicated in colour as follows: 
{\it BeppoSAX} (orange), {\it RXTE} (green), and {\it INTEGRAL} (red). 
 The time of the Chandra observation is shown 
by an arrow. The blue marks in the bottom panel indicate the epochs 
when the \ha\ line displayed an absorption profile.   }
\label{lc_porb}
\end{figure}

\subsection{\emph{Chandra} observations}

Based on the results of our optical monitoring, we triggered an X-ray
observation with \emph{Chandra} on 9 September 2013 (ObsId 15780), when the
\ha\ line was in absorption. A single, uninterrupted 45.4 ks exposure was
carried out. The \emph{Chandra} X-ray Observatory is designed for high
resolution X-ray imaging and spectroscopy in the energy range 0.2-10 keV
 \citep{weisskopf02}. In this work we used data from the Advanced CCD
Imaging Spectrometer (ACIS, \citealt{garmire03}), which provides high
resolution ($\sim$1 arcsec) imaging, and moderate spectral (95 eV at 1.5
keV) and timing reslution ($\sim$ 2.85 ms). For our observation, the source
was placed at the nominal aim point of the back-illuminated ACIS-S3 CCD. To
minimize the possibility of the detector pile up, the observation was
performed in a custom subarray mode with 128 pixel rows, starting from CCD
row 448 which resulted in the CCD frame time of $\sim0.04$\,s. The data
reduction and analysis were performed using the CIAO-4.6.1 analysis package
and the corresponding calibration products (CALDB version 4.6.1.1).

\section{X-ray analysis}

The source counts were extracted from a circular region of radius $\sim5$
arcsec, centered on the source. The extraction radius was chosen to maximise
the signal to noise ratio and encloses $\ge98$\% of the source emission
even at high energies. The background was estimated from two source-free
circular regions with radius of 22 arcseconds adjacent to the source. The
extracted source and background spectra were grouped to contain at least
30\,counts per energy bin in the range from 0.2 to 10\,keV. For the timing
analysis, the photon arrival times were translated to the solar system
barycenter and corrected for motion within the binary system using
ephemeris reported by \cite{baykal07}.

\begin{figure}
\resizebox{\hsize}{!}{\includegraphics{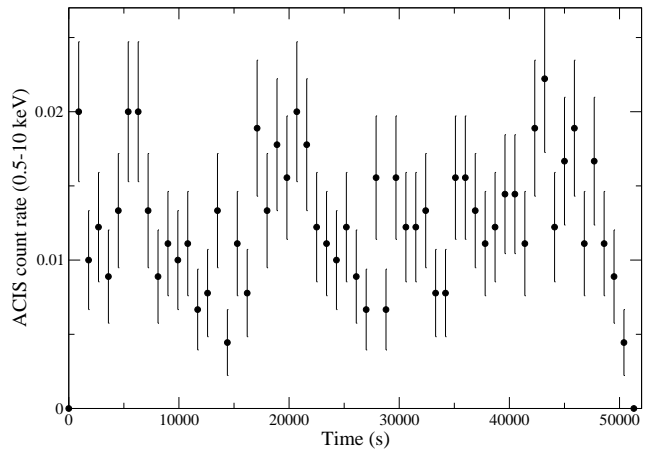} } 
\caption[]{The \emph{Chandra} ACIS light curve binned at a resolution of
900 s in the energy range 0.5--10 keV. Time zero corresponds to MJD 56544.750. }
\label{acis_lc}
\end{figure}

\subsection{Timing analysis}

Figure~\ref{lc_porb} shows the long-term X-ray light curve of \sax,
obtained with the {\it RXTE}/ASM and {\it Swift}/BAT all-sky monitors. The
time of the X-ray observations from various missions is indicated. The ACIS
background-subtracted light curve with bin size of 900 s is shown in
Fig.~\ref{acis_lc}. The average source count rate after background
subtraction is $0.0196\pm0.0006$ counts s$^{-1}$ in the 0.5--10 keV band,
while the background count rate is $0.0051\pm0.0003$ counts s$^{-1}$.

The {\it Chandra} light curve presented above seems to show some
variability on time scales of a few ks with fractional RMS of
$\sim25$\%. However, variability analysis of the event data using the
Gregory-Loredo algorithm (implemented in CIAO tool \emph{glvary})
suggests that the observed emission is consistent with constant rate
(the probability of a variable signal is less than 5\%).

\begin{figure}
\resizebox{\hsize}{!}{\includegraphics{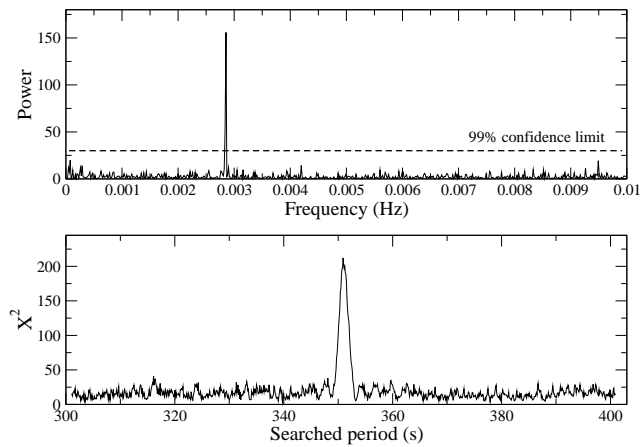} } 
\caption[]{Power spectrum (top) and $\chi^2$ distribution versus folding
period. The dashed line marks the 99\% confidence detection limit. }
\label{spin}
\end{figure}

The power spectrum of the source is also consistent with white noise with
the exception of the strong peak detected at the expected period of \sax\
(see Fig ~\ref{spin}). Indeed, the power spectrum of the 1-10\,keV light
curve with 1\,s time bin size reveals a coherent modulation with maximum
power at $2.853 \times 10^{-3}$ Hz, which corresponds to a pulse period of
$\sim351$ s (Fig.~\ref{spin}). A more accurate determination of the spin
period was obtained through the pulse phase connection technique
\citep{staubert09}. We found the pulse period to be consistent
with constant value of $351.13\pm0.02$\,s (at $1\sigma$ confidence level).

\begin{figure}
\resizebox{\hsize}{!}{\includegraphics{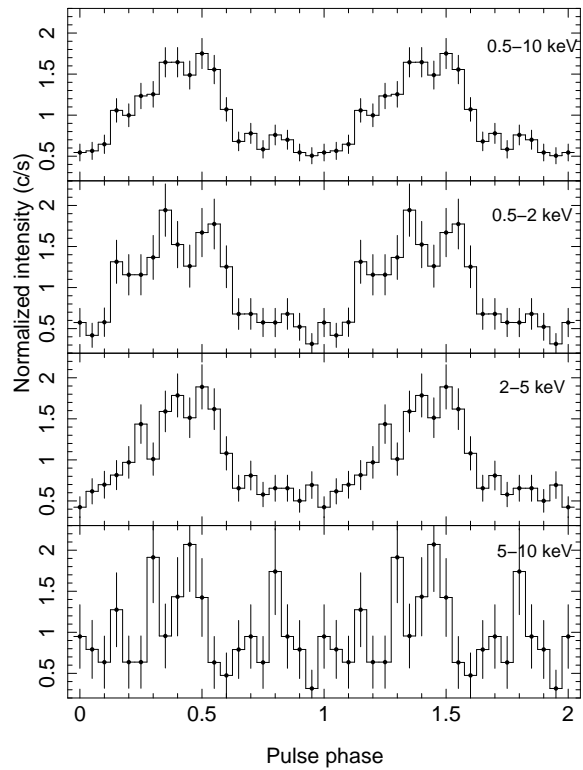} } 
\caption[]{Pulse profile at different energy bands. }
\label{pprofile}
\end{figure}

The background subtracted pulse profiles in several energy ranges folded
with the obtained period are presented in Fig.~\ref{pprofile}. The fraction
of pulsed emission (defined as $PF=(I_{\rm max}-I_{\rm min})/(I_{\rm
max}+I_{\rm min})$, where $I_{\rm min}$ and $I_{\rm max}$, are the minimum
and maximum intensity of the pulse profile) is relatively high (50-80\%)
and independent of energy within the statistical uncertainties. For the
0.5--10 keV band the pulsed fraction is $55\pm8$\%. Similar values have
been reported for \sax\ from {\it XMM-Newton} observations at three orders
of magnitude higher luminosity, although the reported pulse shape was
different \citep[see Fig. 1 in][]{inam04}. The {\it XMM-Newton} profiles
are more asymmetric with a narrow peak followed by a broader one. The
overall {\it Chandra} profiles are more sinusoidal, have only one peak, and
cover a larger fraction of the pulse phase with no sharp features.

\subsection{X-ray spectral analysis}

The source spectrum in the 0.2-10\,keV energy range can be fitted
($\chi^{2}_{\rm red}=1.0$ for 7 degrees of freedom) with a single component
absorbed blackbody. The blackbody temperature $kT=0.98\pm0.07$\,keV, and
radius of $R=0.11\pm0.02 {\rm\,km}$, assuming a distance of 6.5 kpc, are
compatible with the emission from the polar caps of the neutron star. The
absorption column of $N_H=(3\pm1)\times10^{21}{\rm\,atoms\,cm}^{-2}$
(assuming abundances by \citealp{wilms00}) is roughly compatible with
interstellar absorption in the direction of the source
($\sim6\times10^{21}$ \citealt{kalberla05}). The quality of the fit above 6
keV is slightly improved with the inclusion of a power law tail. However,
the limited statistics at high energies makes the parameters of the power
law unconstrained and the overall improvement in terms of $\chi^2$ not
significant.

The 0.5--10 keV unabsorbed X-ray flux is $F_X = 2.3 \times 10^{-13}$ ergs
cm$^{-2}$ s$^{-1}$, which implies an X-ray luminosity of $L_X = 1.2 \times
10^{33}$ ergs s$^{-1}$, assuming a distance of 6.5 kpc \citep{reig04}, or
$L_X = 5.6 \times 10^{32}$ ergs s$^{-1}$, if the distance of 4.5 kpc from
X-ray studies is used \citep{baykal07}. In either case, this is the lowest
luminosity at which X-ray pulsations have been detected in the quiescent
state of an accreting pulsar.

\section{Discussion}

The observational properties and evolution of accreting pulsars are to
a large extent defined by the interaction of the magnetosphere of the
neutron star with the accreting matter. The magnetosphere size is
defined by the magnetic field strength of the neutron star and the
ram pressure of the infalling plasma, and becomes large at low
accretion rates. For a rotating neutron star this implies that at
some point the velocity of the magnetic field lines at the magnetospheric
boundary will exceed local Keplerian velocity and the accretion will
be inhibited \citep{illarionov75, stella86}. The X-ray emission is
expected to switch off at this point. Further interaction
with infalling plasma leads to a spin-down of the neutron star, which
is believed to be the key mechanism behind the relatively long
observed periods of accreting pulsars (compared with expected birth
periods).

Pulsed emission at luminosities $L_X \sim 10^{34}$ erg s$^{-1}$ has been
observed in three systems so far  \citep{rutledge07}. At least in two
occasions, 1A\,0535+262 \citep{negueruela00,doroshenko14} and
IGR\,J21343+4738 \citep{reig14b}, X-ray pulsed emission was detected during
an apparent disk-loss phase. \sax\ is the third case of pulsations detected
during the absence of the Be disk, but it is the first time that X-ray
pulsations are detected at a luminosity level of $\sim 10^{33}$ erg
s$^{-1}$ in the energy range 0.5--10 keV (see Table~\ref{lowlum}).  In what
follows we discuss possible mechanisms that may account for the observed
high-energy emission from \sax.

The minimal accretion luminosity might be found by imposing
$r_m=r_c$, where $r_c$ is the co-rotation radius and $r_m$ the radius
of the magnetosphere \citep[see e.g.][]{campana02}

\begin{eqnarray}
\nonumber
L_{\rm min}(R_{\rm NS}) = 3.9 \times 10^{37} 
k^{7/2}
\left(\frac{B}{10^{12} {\rm \, G}}\right)^2
\left(\frac{P_{\rm spin}}{1 \, {\rm s}}\right)^{-7/3} \\
\left(\frac{M_{\rm X}}{1.4 \, {\rm \msun}}\right)^{-2/3} 
\left(\frac{R_{\rm X}}{10^6 \, {\rm cm}}\right)^5 \, \,  {\rm erg \, s^{-1}} 
\end{eqnarray}

\noindent where $k$ is a constant that accounts for the geometry of the
flow. $k\approx1$ in case of spherical accretion and $k\approx 0.5$ in case
of disk accretion  \citep[see e.g.][and references therein]{ikhsanov10}. 
$B$ is the magnetic field strength, $P_{\rm spin}$ the spin period, and
$M_{\rm X}$ and $R_{\rm X}$ the mass and radius of the neutron star,
respectively.

The long spin period of \sax\ and the detection of X-ray pulsations put
this source along with 1A 0535+26, 4U 1145--619, and 1A 1118--615, in a
category of systems with accretion powered quiescent emission
\citep{rutledge07,doroshenko14}. Unfortunately, there is no direct estimate
of the magnetic field of the neutron star in \sax, so it is not clear
whether it is in the centrifugally inhibited propeller state. Nevertheless,
the fact that pulsed emission is detected suggests that this is likely not
the case. We can turn the argument around and derive an upper limit on the
magnetic field assuming that the source does not enter the centrifugally
inhibited state. For a source distance of 6.5 kpc, $P_{\rm spin}=351$ s,
$k=1$, and $L_{\rm min}\approx L_{\rm obs}\approx 1.2 \times 10^{33}$ erg
s$^{-1}$, the magnetic field must be $\simless 5.2\times 10^{12}$ G. For
higher magnetic fields, $L_{\rm min} > L_{\rm obs}$ and the accreting
matter would no longer reach the neutron star surface because it would be
spun away by the fast rotation of the magnetosphere.

It is interesting to compare this value with estimates of the magnetic
field from the spin evolution history of the pulsar. Based on the
correlation of the accreting luminosity and the spin-up rate and the
\citet{ghosh79} model, various authors \citep{baykal02,baykal07,ducci08}
have estimated the magnetic field to be $B\sim 1-3 \times 10^{13}$ G, i.e.
the accretion should be inhibited. On the other hand,  \citet{sidoli05}
obtained $B\sim 10^{12}$ G. The uncertainty mainly stems from the distance,
which is also a parameter obtained from the fit to the pulse period
derivative-luminosity correlation. A magnetic field strength above
$10^{13}$ G requires a distance to the source $<5$ kpc. If the distance of
6.5 kpc obtained from optical observations \citep{reig04} is considered,
then $B\sim 10^{12}$ G \citep{sidoli05}, and the observed X-ray emission
might be powered by accretion. In the absence of the Be disk, accretion
likely proceeds directly from the stellar wind of the companion.
Alternatively, the so-called ``dead accretion disks'' proposed by
\citet{syunyaev77} might represent another source of matter for accretion.

\begin{figure}
\resizebox{\hsize}{!}{\includegraphics{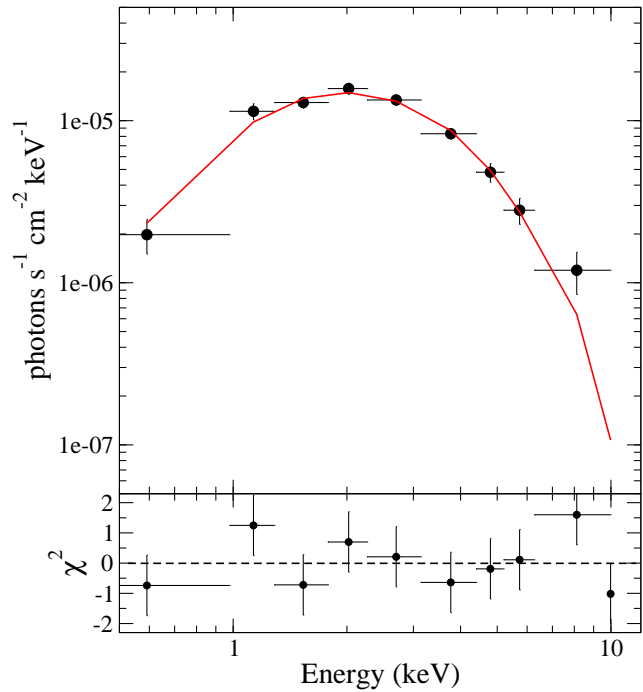} } 
\caption[]{{\it Chandra}/ACIS-S3 spectrum of \sax. The solid line is the
best-fitting model composed of an absorbed  blackbody.}
\label{xsp}
\end{figure}

\begin{table*}
\begin{center}
\caption{Detection of X-rays from Be accreting pulsars at low luminosities.}
\label{lowlum}
\begin{tabular}{lcccclll}
\hline \hline \noalign{\smallskip}
Source		&$P_{\rm spin}$	 &$L_X$			&Energy range   &Mission		&Distance	&Equatorial	&Propoller\\
		&(s)		&(erg s$^{-1}$)		&(keV)	    	&			&(kpc)		&disk		&mechanism \\
\hline \noalign{\smallskip}
\multicolumn{8}{c}{Pulsed emission detected}\\
\hline \noalign{\smallskip}
SAX\,J2103.5+4545&$351.03\pm0.05$ &$1.2\times10^{33}$	&0.5--10  	&{\it Chandra}$^1$	&6.5$^a$	&no	&no?  \\
1A\,0535+26	&$103.5$	 &$3.5\times10^{33}$	&3--20  	&{\it RXTE}$^2$		&2$^b$		&no	&yes  \\
		&$103.41\pm0.02$ &$1.5\times10^{33}$	&2--10  	&{\it BeppoSAX}$^{3,4}$	&2$^b$		&yes	&yes? \\
		&$103.286\pm0.006$ &$1.3\times10^{34}$	&0.2--12	&{\it XMM-Newton}$^5$	&2$^b$		&yes	&yes? \\
4U\,1145--619	&$290\pm2$	 &$5.9\times10^{33}$	&0.5--2 	&{\it Einstein}$^6$	&3.1$^c$	&yes?	&no   \\
1A\,1118--615	&$409.2\pm0.2$	 &$1.8\times10^{33}$	&0.5--10	&{\it Chandra}$^7$	&5$^d$		&yes	&no \\	
\hline \noalign{\smallskip}
\multicolumn{8}{c}{Pulsed emission not detected}\\
\hline \noalign{\smallskip}
Cep X--4	&--		&$3.2\times10^{32}$	&0.1--2.5	&{\it ROSAT}$^8$		&3.8$^e$	&?	&yes\\
4U\,0115+63	&--		&$8.4\times10^{32}$	&0.5--10	&{\it BeppoSAX}$^9$	&8$^f$		&yes	&yes\\	
V 0332+53	&--		&$5.3\times10^{32}$	&0.5--10	&{\it Chandra}$^9$	&7$^g$		&yes?	&yes\\
IGR\,J01363+6610 &--		&$9.1\times10^{31}$	&0.2--12	&{\it XMM-Newton}$^{10}$&2$^h$		&yes	&?\\
GRO\,J2058+42	&--		&$5.6\times10^{33}$	&1--10		&{\it Chandra}$^{11}$	&9$^i$		&yes?	&?\\
\hline \hline \noalign{\smallskip}
\end{tabular}
\end{center}
\medskip
$^1$:this work      ; $^2$:\citet{negueruela00} ; $^3$:\citet{orlandini04}; $^4$:\citet{mukherjee05} ;
$^5$: \citet{doroshenko14} ; $^6$:\citet{mereghetti87} ; $^7$:\citet{rutledge07} ; $^8$: \citet{schulz95} ; 
$^9$:\citet{campana02} ; $^{10}$:\citet{tomsick11} ; $^{11}$:\citet{wilson05} ;
$^a$:\citet{reig04} ; $^b$:\citet{steele98}     ; $^c$:\citet{stevens97} ; $^d$:\citet{janot81} ; 
$^e$:\citet{bonnet98} ; $^f$: \citet{reig07b}  ; $^g$:\citet{negueruela99} : $^h$:\citet{reig05a} ;
$^i$:\citet{wilson05}
\end{table*}

The main problem with the accretion interpretation is that the X-ray
emission is expected to be rather variable regardless of the accretion
mechanism, whereas \sax\ exhibits no detectable variability in the
\emph{Chandra} observation. Indeed, all other sources where pulsations have
been detected in quiescence (i.e. 4U 1145--619, 1A 1118--615, 1A 0535+26)
exhibit strong low frequency noise usually associated with accretion. In
fact, \sax\ exhibits this type of variability as well at higher
luminosities \citep{inam04}. In contrast, the \emph{Chandra} power spectrum
is consistent with white noise with the exception of the pulsation peak.

\begin{figure}[t]
	\centering
		\includegraphics[width=8cm,height=6cm]{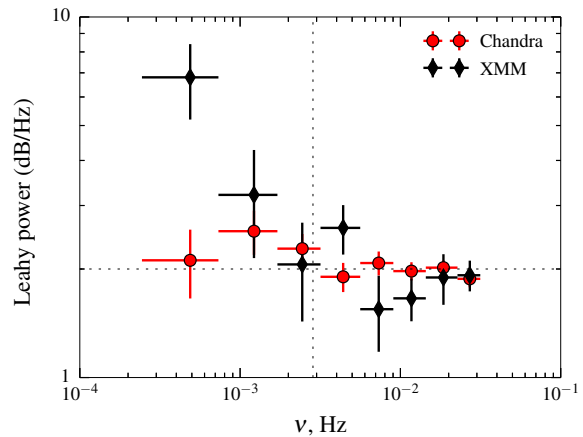}
	\caption{Comparison of the noise power spectra of \sax\ at high
	(\emph{XMM-Newton}) and low (\emph{Chandra}) luminosities. 
	The power is normalized 	so that the white noise level has 
	a power of two \citep{leahy83}, i.e. the \emph{Chandra} lightcurve is 
	consistent with white noise.
	Counting statistics of the \emph{XMM-Newton}
	observation was adjusted to match that of \emph{Chandra} as described
	in the text. Note the absence of the low frequency noise component 
	in \emph{Chandra} power spectrum.
	}
	\label{fig:psd}
\end{figure}

Of course, with $\sim900$ photons detected from the source this might be
simply due to the insufficient statistics. To clarify whether this is the
case, we have compared our \emph{Chandra} results with the
\emph{XMM-Newton} observation of \sax\ at higher flux level (obsid
0149550401) where the statistics is much better and low frequency noise is
apparent. From more than $2\times10^{5}$ source photons detected by EPIC PN
camera, we randomly selected 900 to match the \emph{Chandra} statistics,
and investigated the power spectrum of the resulting light curves. In both
cases X-ray pulsations are clearly detected and dominate the power
spectrum. To investigate the noise properties we had, therefore, to
subtract the pulsed flux from the light curve prior to binning of the power
spectrum \citep{revnivtsev09,doroshenko14}.  In particular, a synthetic
lightcurve containing a repeated observation-averaged pulse profile was
subtracted from the observed lightcurve, which was sufficient to suppress
the observed pulsations below detectability with available statistics. The
results are presented in Fig.~\ref{fig:psd}. The excess of power at low
frequencies remains apparent in the \emph{XMM-Newton} data set even with
the reduced statistics, but it is completely absent in the \emph{Chandra}
light curve despite comparable counting statistics (we ignore the
background in both cases, which is slightly higher in the \emph{XMM-Newton}
observation, so \emph{Chandra} should in fact be even more sensitive to the
presence of low frequency noise from the source). We argue, therefore, that
the observed change in the power spectrum is likely real and might imply
that accretion is not responsible for the observed \emph{Chandra} flux.

Another scenario that has been put forward to explain the quiescent
emission of accreting neutron star is deep crustal heating
\citep{brown98,wijnands13}. During the accretion phase, the crust of a
neutron star is heated by nuclear reactions (mainly by beta captures and
pycnonuclear reactions). This heat is conducted inwards, heating the core,
and outwards, where it is emitted as thermal emission from the surface.
After the accreting active period, the crust of a neutron star cools by
X-ray emission until it reaches thermal equilibrium with the core emission
corresponding to the quiescent state. The X-ray luminosity in this state
depends on the time-averaged accretion rate as $L_q\sim 6.03 \times
10^{32}(\dot{M}/1\times10^{-11}$ $\msun$ yr$^{-1}$), where $\dot{M}$ is the
average accretion rate including outbursts \citep{brown98}.

To estimate $\dot{M}$ we used the 70-month average {\it Swift}/BAT spectrum
fitting it with a cutoff power law in the 15-100 keV band. The derived
source flux is be $3.6\times 10^{-11}$ erg cm$^{-2}$ s$^{-1}$. The average
count rate obtained from the 12 year {\it RXTE}/ASM light curve (see
Fig.~\ref{lc_porb}) is 0.3 count s$^{-1}$ or 4 mcrab, which corresponds to
a flux of $9.6\times 10^{-11}$ erg cm$^{-2}$ s$^{-1}$ in the 2--10 keV
band. The overall 1--100 keV luminosity assuming a distance of 6.5 kpc is
then $6.5\times 10^{35}$ erg s$^{-1}$ and the mass accretion rate
$\dot{M}=5.6\times 10^{-11}$ $\msun$ yr$^{-1}$. Here we assumed the
canonical mass and radius of a neutron star and maximum efficiency in the
conversion of luminosity into mass accretion rate. Thus the time average
luminosity in quiescence expected by deep crustal heating is $L_q\sim
3\times 10^{33}$ erg s$^{-1}$, which roughly agrees with the observed value.
The problem with the incandescent neutron star scenario is that the
blackbody temperature is too high, and the size of the emitting region is
too small to be the entire surface of the neutron star \citep[but see
discussion in][]{brown98}.

The relatively high temperature thermal spectrum, in combination with the
large pulse fraction and broad pulse profiles indicate that the emission
arises from a rotating region which is hotter (and more luminous) than the
rest of the surface of the pulsar.  Thus non-uniform cooling, primarily
through the polar regions, has to be invoked for this scenario to remain
valid.  The most obvious explanation is that the observed emission comes
from the polar cap, which could be heated by sporadic accretion. Note that
our \emph{Chandra} observation took place in between two X-ray outbursts
(Fig.\ref{lc_porb}). In this situation, the polar caps does not have the
time to cool down.

\section{Conclusions}

We have analysed the optical and X-ray emission of the Be/X-ray binary
\sax\ in deep X-ray quiescence. The optical spectra indicate that the Be
star's circumstellar disk was absent during the X-ray observation. The
X-ray luminosity is the lowest so far observed in this source (more than
three orders of magnitude lower than in the bright state) and constitutes
the lowest luminosity level for which X-ray pulsations have been detected
in an accreting pulsar. The absence of any variability typical for
accreting neutron star suggests that the observed emission likely
originates from the polar caps of the neutron stars heated during
intermittent accretion episodes or by non-uniform cooling of the neutron
star after a recent outburst.

\section*{Acknowledgments}

VD thank the Deutsches Zentrums für Luft- und Raumfahrt (DLR) and
Deutsche Forschungsgemein- schaft (DFG) for financial support (grant  DLR
50 OR 0702). This research has made use of software provided by the Chandra
X-ray Center (CXC) in the application packages CIAO. This work has made use
of NASA's Astrophysics Data System Bibliographic Services and of the SIMBAD
database, operated at the CDS, Strasbourg, France. Skinakas Observatory is
run by the University of Crete and the Foundation for Research and
Technology-Hellas.


\end{document}